\documentclass{article}%
\usepackage{amsmath}
\usepackage{amsfonts}
\usepackage{amssymb}
\usepackage{graphicx}%
\newtheorem{theorem}{Theorem}

\newtheorem{corollary}[theorem]{Corollary}

\newtheorem{proposition}[theorem]{Proposition}
\newtheorem{remark}[theorem]{Remark}

\newcommand{\bpartial}{\mathop{\partial\kern -4pt\raisebox{.8pt}{$|$}}}
\newcommand{\bra}{\mathopen{[\kern-1.6pt[}}
\newcommand{\ket}{\mathclose{]\kern-1.5pt]}}
\newcommand{\bbra}{\mathopen{[\kern-2.2pt[\kern-2.3pt[}}
\newcommand{\bket}{\mathclose{]\kern-2.1pt]\kern-2.3pt]}}
\newcommand{\slg}{\mbox{\bfseries\slshape g}}

\makeindex
\begin{document}

\title{Killing Vector Fields, Maxwell Equations and Lorentzian
Spacetimes\thanks{Paper presented at the 8$^{th}$ International Conference on
Clifford Algebras and their Applications in Mathematical Physics (ICCA8),
Campinas, May 26-30 2008.}}
\author{\hspace{-1cm}Waldyr A. Rodrigues Jr.\\$\hspace{-0.1cm}$Institute of Mathematics, Statistics and Scientific Computation\\IMECC-UNICAMP CP 6065\\13083-859 Campinas, SP, Brazil\\e-mail: walrod@ime.unicamp.br}
\maketitle

\begin{abstract}
In this paper we first analyze the structure of Maxwell equations in a
Lorentzian spacetime when the potential is $A=\mathfrak{e}K$, i.e.,
proportional to a $1$-form $K$ physically equivalent Killing vector field. We
show that $A$ necessarily obeys the Lorenz gauge $\delta A=0$. Moreover we
determine the form of the current associated with this potential showing that
it is of a superconducting type, i.e., \ proportional to the potential and
given by $2\mathcal{R}_{\beta}A^{\beta}$, where the $\mathcal{R}_{\beta}$ are
the Ricci $1$-form fields. Finally we study the structure of the spacetime
generated by the coupled system consisting of a electromagnetic field $F=dA$
(with $A=\mathfrak{e}K)$, an ideal charged fluid with dynamics described by an
action function $S$ and the gravitational field. We show that Einstein
equations is then equivalent to Maxwell equations with a current \ given by
$fFAF$ (the product meaning the Clifford product of the corresponding fields),
where $f$ is a scalar function which satisfies a well determined algebraic
quadratic equation.

\end{abstract}

\section{Introduction}

In a previous paper \cite{notte} \ we study using the Clifford bundle
formalism the effective\ Lorentzian and teleparallel spacetimes generated by a
electromagnetic field moving in Minkowski spacetime.

Here, using the same mathematical apparatus, we study another intriguing
connections between gravitation described by Einstein field equations and
electromagnetism described by Maxwell equations. In order to do that we first
prove in Section 2 a proposition showing that if $\mathbf{K}$ is a Killing
vector field on a Lorentzian manifold $(M,%
%TCIMACRO{\TeXButton{slg}{\slg}}%
%BeginExpansion
\slg
%EndExpansion
)$ then the form field $K=$ $%
%TCIMACRO{\TeXButton{slg}{\slg}}%
%BeginExpansion
\slg
%EndExpansion
(\mathbf{K},$ $)$ satisfies $\delta K=0$ and a wave equation given by
Eq.(\ref{t2}) in terms of the covariant D' Alembertian. We also show that the
Ricci operator (which can be defined only in the Clifford bundle of
differential forms) applied to $K$ it is equal to the covariant D'Alembertian
\ applied to $K$. Next, in Section 3 we analyze the structure of Maxwell
equations in a Lorentzian spacetime when the potential obeys the Lorenz gauge.
Take notice that if a potential is in Lorenz gauge this does not necessarily
implies that it is a $1$-form physically equivalent to a Killing vector field.
Moreover we determine the form of the current associated with this potential
$A$ showing that it is given by\footnote{Our result differs from a factor of
$2$ from the one presented in \cite{papapetrou} and also in \cite{fayos},
where an electromagnetic potential proportional to a Killing vector field is
called a \textit{Papapetrou field}. The important discrepancy is due to the
fact that those authors identified the electromagnetic current $J_{e}$ with
$\square A$ instead of identifying it with $\ -\delta dA$, as it must be. See
the text for details.} $2A^{\beta}\mathcal{R}_{\beta}$, where the
$\mathcal{R}_{\beta}$ are the Ricci $1$-form fields (Eq.(\ref{166})) In
Section 4 we study the structure of the Lorentzian spacetime representing the
gravitational field produced and interacting with an electromagnetic field
$F=dA$ (where $A$ is proportional $K=$ $%
%TCIMACRO{\TeXButton{slg}{\slg}}%
%BeginExpansion
\slg
%EndExpansion
(\mathbf{K},$ $)$, with $\mathbf{K}$ a Killing vector field) generated by an
\textit{ideal} charged current $J_{e}$. We show that Einstein equations is in
this case represented by Maxwell equations with a current given by $fFAF$ (the
product being intended as the Clifford product of the corresponding fields),
where $f$ is a scalar function solution of a well determined algebraic
quadratic equation (Eq.(\ref{cs17}). In Section 5 we present our conclusions
and in\ the Appendix we recall the main definitions and formulas of the
Clifford bundle formalism, proving a result that is need in the proof of
Proposition 1.

\section{Some Preliminaries}

In this paper a spacetime structure is a pentuple $\mathfrak{M}=(M,%
%TCIMACRO{\TeXButton{slg}{\slg}}%
%BeginExpansion
\slg
%EndExpansion
,D,\tau_{g},\uparrow)$\ where $(M,%
%TCIMACRO{\TeXButton{slg}{\slg}}%
%BeginExpansion
\slg
%EndExpansion
,\tau_{g})$ is a Lorentzian manifold, $D$ is the Levi-Civita connection of $%
%TCIMACRO{\TeXButton{slg}{\slg}}%
%BeginExpansion
\slg
%EndExpansion
$ and $\uparrow$ is an equivalence relation between timelike vector fields
defining the time orientation\footnote{Details may be found, e.g., in
\cite{rodoliv2007,sawu}}. Also, $\mathtt{g}\in\sec T_{2}^{0}M$ denotes the
metric of the cotangent bundle, $\bigwedge T^{\ast}M$ denotes the bundle of
(nonhomogeneous) differential forms and $\mathcal{C\ell(}M,\mathtt{g})$
denotes the Clifford bundle of differential forms. We shall take advantage of
the well known fact that \cite{lawmi} $\bigwedge T^{\ast}M\hookrightarrow
\mathcal{C\ell(}M,\mathtt{g})$ and use in our calculations the powerful
Clifford bundle formalism \cite{rodoliv2007}. Let $\{\partial_{\mu}\}$,
$\partial_{\mu}:=\frac{\partial}{\partial x^{\mu}}$ be an arbitrary coordinate
basis for $TU\subset TM$ and $\{%
%TCIMACRO{\TeXButton{bg}{\mbox{\boldmath{$\gamma$}}}}%
%BeginExpansion
\mbox{\boldmath{$\gamma$}}%
%EndExpansion
^{\mu}=dx^{\mu}\}$ the corresponding dual basis of $T^{\ast}U\subset T^{\ast
}M=\bigwedge^{1}T^{\ast}M$. As explained in the Appendix the $%
%TCIMACRO{\TeXButton{bg}{\mbox{\boldmath{$\gamma$}}}}%
%BeginExpansion
\mbox{\boldmath{$\gamma$}}%
%EndExpansion
^{\mu}$ will be though as sections of the Clifford bundle, more precisely, $%
%TCIMACRO{\TeXButton{bg}{\mbox{\boldmath{$\gamma$}}}}%
%BeginExpansion
\mbox{\boldmath{$\gamma$}}%
%EndExpansion
^{\mu}\in\sec T^{\ast}U\subset\sec\bigwedge^{1}T^{\ast}M\hookrightarrow
\mathcal{C\ell(}M,\mathtt{g})$. Also, we recall that the set $\{\partial^{\mu
}\}$, $\partial^{\mu}=g^{\mu\nu}\frac{\partial}{\partial x^{\mu}}\in\sec TM$
such that $%
%TCIMACRO{\TeXButton{slg}{\slg}}%
%BeginExpansion
\slg
%EndExpansion
(\partial_{\mu},\partial^{\nu})=\delta_{\nu}^{\mu}$ is called the reciprocal
basis of $\{\partial_{\mu}\}$ and the set $\{%
%TCIMACRO{\TeXButton{bg}{\mbox{\boldmath{$\gamma$}}}}%
%BeginExpansion
\mbox{\boldmath{$\gamma$}}%
%EndExpansion
_{\mu}\}$ such that \texttt{g}$(%
%TCIMACRO{\TeXButton{bg}{\mbox{\boldmath{$\gamma$}}}}%
%BeginExpansion
\mbox{\boldmath{$\gamma$}}%
%EndExpansion
^{\mu},%
%TCIMACRO{\TeXButton{bg}{\mbox{\boldmath{$\gamma$}}}}%
%BeginExpansion
\mbox{\boldmath{$\gamma$}}%
%EndExpansion
_{\nu})=\delta_{\nu}^{\mu}$ is called the reciprocal basis of $\{%
%TCIMACRO{\TeXButton{bg}{\mbox{\boldmath{$\gamma$}}}}%
%BeginExpansion
\mbox{\boldmath{$\gamma$}}%
%EndExpansion
^{\mu}\}$. We denote \texttt{g}$(%
%TCIMACRO{\TeXButton{bg}{\mbox{\boldmath{$\gamma$}}}}%
%BeginExpansion
\mbox{\boldmath{$\gamma$}}%
%EndExpansion
^{\mu},%
%TCIMACRO{\TeXButton{bg}{\mbox{\boldmath{$\gamma$}}}}%
%BeginExpansion
\mbox{\boldmath{$\gamma$}}%
%EndExpansion
_{\nu})=%
%TCIMACRO{\TeXButton{bg}{\mbox{\boldmath{$\gamma$}}}}%
%BeginExpansion
\mbox{\boldmath{$\gamma$}}%
%EndExpansion
^{\mu}\cdot%
%TCIMACRO{\TeXButton{bg}{\mbox{\boldmath{$\gamma$}}}}%
%BeginExpansion
\mbox{\boldmath{$\gamma$}}%
%EndExpansion
_{\nu}$, where $\cdot$ denotes the scalar product in $\mathcal{C\ell
(}M,\mathtt{g})$.\ Finally, ${%
%TCIMACRO{\TeXButton{dirac}{\mbox{\boldmath$\partial$}}}%
%BeginExpansion
\mbox{\boldmath$\partial$}%
%EndExpansion
=}%
%TCIMACRO{\TeXButton{bg}{\mbox{\boldmath{$\gamma$}}}}%
%BeginExpansion
\mbox{\boldmath{$\gamma$}}%
%EndExpansion
^{\mu}D_{\partial_{\mu}}$ denotes the Dirac operator acting on sections of
$\mathcal{C\ell(}M,\mathtt{g})$ and $\mathcal{\square=}%
%TCIMACRO{\TeXButton{dirac}{\mbox{\boldmath$\partial$}}}%
%BeginExpansion
\mbox{\boldmath$\partial$}%
%EndExpansion
{\cdot%
%TCIMACRO{\TeXButton{dirac}{\mbox{\boldmath$\partial$}}}%
%BeginExpansion
\mbox{\boldmath$\partial$}%
%EndExpansion
}$ and ${%
%TCIMACRO{\TeXButton{dirac}{\mbox{\boldmath$\partial$}}}%
%BeginExpansion
\mbox{\boldmath$\partial$}%
%EndExpansion
\wedge%
%TCIMACRO{\TeXButton{dirac}{\mbox{\boldmath$\partial$}}}%
%BeginExpansion
\mbox{\boldmath$\partial$}%
%EndExpansion
}$ denotes respectively the covariant D'Alembertian and the Ricci operators.
The operator $\Diamond={%
%TCIMACRO{\TeXButton{dirac}{\mbox{\boldmath$\partial$}}}%
%BeginExpansion
\mbox{\boldmath$\partial$}%
%EndExpansion
}^{2}$ is called Hodge D'Alembertian and the relations between those operators
and their main properties are presented in the Appendix.

\begin{proposition}
Let $\mathbf{K}\in\sec TM$ be a Killing vector field, i.e.,
$\pounds _{\mathbf{K}}%
%TCIMACRO{\TeXButton{slg}{\slg}}%
%BeginExpansion
\slg
%EndExpansion
=0$. Let $K=%
%TCIMACRO{\TeXButton{slg}{\slg}}%
%BeginExpansion
\slg
%EndExpansion
(\mathbf{K},)$. Then
\begin{align}
\text{\emph{ }}\delta K  &  =0,\label{t1}\\
{%
%TCIMACRO{\TeXButton{dirac}{\mbox{\boldmath$\partial$}}}%
%BeginExpansion
\mbox{\boldmath$\partial$}%
%EndExpansion
\wedge%
%TCIMACRO{\TeXButton{dirac}{\mbox{\boldmath$\partial$}}}%
%BeginExpansion
\mbox{\boldmath$\partial$}%
%EndExpansion
}K  &  =K_{\alpha}\mathcal{R}^{\alpha},\label{t2}\\
\text{\emph{ }}\square K  &  =K_{\alpha}\mathcal{R}^{\alpha}, \label{t3}%
\end{align}
where $\mathcal{R}^{\alpha}\in\sec\bigwedge^{1}T^{\ast}M\hookrightarrow
\mathcal{C\ell(}M,\mathtt{g})$ are the Ricci $1$-forms given by
\emph{Eq.(\ref{166}).}
\end{proposition}

\noindent\textbf{Proof \ }To prove Eq.(\ref{t1}) it is only necessary to
recall that since
\begin{equation}
\pounds _{\mathbf{K}}%
%TCIMACRO{\TeXButton{slg}{\slg}}%
%BeginExpansion
\slg
%EndExpansion
=0\Leftrightarrow D_{\mu}K_{\nu}+D_{\nu}K_{\mu}=0 \label{kil}%
\end{equation}
and by Eq.(\ref{13}) $\delta K=-{%
%TCIMACRO{\TeXButton{dirac}{\mbox{\boldmath$\partial$}}}%
%BeginExpansion
\mbox{\boldmath$\partial$}%
%EndExpansion
}{\lrcorner}K$ we have%
\begin{align*}
\delta K  &  =-%
%TCIMACRO{\TeXButton{bg}{\mbox{\boldmath{$\gamma$}}}}%
%BeginExpansion
\mbox{\boldmath{$\gamma$}}%
%EndExpansion
^{\mu}{\lrcorner}D_{e_{\mu}}K\\
&  =-%
%TCIMACRO{\TeXButton{bg}{\mbox{\boldmath{$\gamma$}}}}%
%BeginExpansion
\mbox{\boldmath{$\gamma$}}%
%EndExpansion
^{\mu}{\lrcorner\lbrack(}D_{_{\mu}}K_{\nu})%
%TCIMACRO{\TeXButton{bg}{\mbox{\boldmath{$\gamma$}}}}%
%BeginExpansion
\mbox{\boldmath{$\gamma$}}%
%EndExpansion
^{\nu}]\\
&  =g^{\mu\nu}D_{_{\mu}}K_{\nu}=\frac{1}{2}g^{\mu\nu}(D_{_{\mu}}K_{\nu
}+D_{_{\nu}}K_{\mu})=0.
\end{align*}

The proof of Eq.(\ref{t2}) is trivial. Indeed, the Ricci operator is
extensorial \cite{rodoliv2007}, i.e., according to Eq.(\ref{16a}) satisfies \
\[
{%
%TCIMACRO{\TeXButton{dirac}{\mbox{\boldmath$\partial$}}}%
%BeginExpansion
\mbox{\boldmath$\partial$}%
%EndExpansion
\wedge%
%TCIMACRO{\TeXButton{dirac}{\mbox{\boldmath$\partial$}}}%
%BeginExpansion
\mbox{\boldmath$\partial$}%
%EndExpansion
}K=K_{\mu}{%
%TCIMACRO{\TeXButton{dirac}{\mbox{\boldmath$\partial$}}}%
%BeginExpansion
\mbox{\boldmath$\partial$}%
%EndExpansion
\wedge%
%TCIMACRO{\TeXButton{dirac}{\mbox{\boldmath$\partial$}}}%
%BeginExpansion
\mbox{\boldmath$\partial$}%
%EndExpansion
}%
%TCIMACRO{\TeXButton{bg}{\mbox{\boldmath{$\gamma$}}}}%
%BeginExpansion
\mbox{\boldmath{$\gamma$}}%
%EndExpansion
^{\mu}%
\]
and thus using Eq.(\ref{166}) we get:
\begin{equation}
{%
%TCIMACRO{\TeXButton{dirac}{\mbox{\boldmath$\partial$}}}%
%BeginExpansion
\mbox{\boldmath$\partial$}%
%EndExpansion
\wedge%
%TCIMACRO{\TeXButton{dirac}{\mbox{\boldmath$\partial$}}}%
%BeginExpansion
\mbox{\boldmath$\partial$}%
%EndExpansion
}K=\mathcal{R}_{\mu}K^{\mu}. \label{dd}%
\end{equation}

To prove Eq.(\ref{t3}) we use Eq.(\ref{15}) and write%
\begin{equation}
{%
%TCIMACRO{\TeXButton{dirac}{\mbox{\boldmath$\partial$}}}%
%BeginExpansion
\mbox{\boldmath$\partial$}%
%EndExpansion
\cdot%
%TCIMACRO{\TeXButton{dirac}{\mbox{\boldmath$\partial$}}}%
%BeginExpansion
\mbox{\boldmath$\partial$}%
%EndExpansion
}K=g^{\sigma\nu}D_{\sigma}D_{\nu}K_{\mu}%
%TCIMACRO{\TeXButton{bg}{\mbox{\boldmath{$\gamma$}}}}%
%BeginExpansion
\mbox{\boldmath{$\gamma$}}%
%EndExpansion
^{\mu} \label{p31}%
\end{equation}

Now, \ we calculate $D_{\sigma}D_{\nu}K_{\alpha}$. Since $\mathbf{K}$ is a
Killing vector fied satisfying Eq.(\ref{kil}) we can write
\begin{align}
&  D_{\sigma}(D_{\nu}K_{\mu}+D_{\mu}K_{\nu})\nonumber\\
&  =[D_{\sigma},D_{\nu}]K_{\mu}+D_{\nu}D_{\sigma}K_{\mu}+[D_{\sigma},D_{\mu
}]K_{\nu}+D_{\mu}D_{\sigma}K_{\nu}=0. \label{p32}%
\end{align}
Taking into account that%
\begin{align}
g^{\sigma\nu}[D_{\sigma},D_{\nu}]K_{\mu}  &  =0,\nonumber\\
g^{\sigma\nu}D_{\mu}D_{\sigma}K_{\nu}  &  =\frac{1}{2}g^{\sigma\nu}D_{\mu
}(D_{\sigma}K_{\nu}+D_{\nu}K_{\sigma})=0,\nonumber\\
g^{\sigma\nu}[D_{\sigma},D_{\mu}]K_{\nu}  &  =-g^{\sigma\nu}R_{\nu
\;\ \sigma\mu}^{\;\text{\ }\rho}K_{\rho}=-g^{\sigma\nu}R_{\nu\rho\sigma\mu
}K^{\rho}\nonumber\\
&  =-g^{\sigma\nu}R_{\rho\nu\mu\sigma}K^{\rho}=-R_{\rho\mu}K^{\rho},
\label{p33}%
\end{align}
we get on multiplying Eq.(\ref{p32}) by $g^{\sigma\nu}$ that
\[
g^{\sigma\nu}D_{\nu}D_{\sigma}K_{\mu}=R_{\rho\mu}K^{\rho},
\]
and thus
\[
{%
%TCIMACRO{\TeXButton{dirac}{\mbox{\boldmath$\partial$}}}%
%BeginExpansion
\mbox{\boldmath$\partial$}%
%EndExpansion
\cdot%
%TCIMACRO{\TeXButton{dirac}{\mbox{\boldmath$\partial$}}}%
%BeginExpansion
\mbox{\boldmath$\partial$}%
%EndExpansion
}K=g^{\sigma\nu}D_{\sigma}D_{\nu}K_{\mu}%
%TCIMACRO{\TeXButton{bg}{\mbox{\boldmath{$\gamma$}}}}%
%BeginExpansion
\mbox{\boldmath{$\gamma$}}%
%EndExpansion
^{\mu}=R_{\rho\mu}K^{\rho}%
%TCIMACRO{\TeXButton{bg}{\mbox{\boldmath{$\gamma$}}}}%
%BeginExpansion
\mbox{\boldmath{$\gamma$}}%
%EndExpansion
^{\mu}=K^{\rho}\mathcal{R}_{\alpha}.
\]

\begin{corollary}
Call $M=dK$. Then
\begin{equation}
J=:-\delta M=2K^{\beta}\mathcal{R}_{\beta} \label{t14}%
\end{equation}

\end{corollary}

\noindent\textbf{Proof }\ Indeed, we have recalling Eq.(\ref{t1}) and
Eq.(\ref{13bis}) that
\begin{align}
-\delta dK  &  =-\delta dK-d\delta K=(d-\delta)(d-\delta)K={%
%TCIMACRO{\TeXButton{dirac}{\mbox{\boldmath$\partial$}}}%
%BeginExpansion
\mbox{\boldmath$\partial$}%
%EndExpansion
}^{2}K\nonumber\\
&  ={%
%TCIMACRO{\TeXButton{dirac}{\mbox{\boldmath$\partial$}}}%
%BeginExpansion
\mbox{\boldmath$\partial$}%
%EndExpansion
\cdot%
%TCIMACRO{\TeXButton{dirac}{\mbox{\boldmath$\partial$}}}%
%BeginExpansion
\mbox{\boldmath$\partial$}%
%EndExpansion
}K+{%
%TCIMACRO{\TeXButton{dirac}{\mbox{\boldmath$\partial$}}}%
%BeginExpansion
\mbox{\boldmath$\partial$}%
%EndExpansion
\wedge%
%TCIMACRO{\TeXButton{dirac}{\mbox{\boldmath$\partial$}}}%
%BeginExpansion
\mbox{\boldmath$\partial$}%
%EndExpansion
}K\nonumber\\
&  =K^{\beta}\mathcal{R}_{\beta}+K^{\beta}\mathcal{R}_{\beta}=2K^{\beta
}\mathcal{R}_{\beta}, \label{t12}%
\end{align}
and the result is proved.

\section{Electrodynamics on $\mathfrak{M}$}

As it is well known in General Relativity (GR) the gravitational field
generated by an energy momentum tensor $\mathbf{T=}T_{\mu}\otimes$ $%
%TCIMACRO{\TeXButton{bg}{\mbox{\boldmath{$\gamma$}}}}%
%BeginExpansion
\mbox{\boldmath{$\gamma$}}%
%EndExpansion
^{\mu}$, (where the $T_{\mu}=T_{\mu}^{\nu}%
%TCIMACRO{\TeXButton{bg}{\mbox{\boldmath{$\gamma$}}}}%
%BeginExpansion
\mbox{\boldmath{$\gamma$}}%
%EndExpansion
_{\nu}\in\sec\bigwedge^{1}T^{\ast}M\hookrightarrow\mathcal{C\ell(}%
M,\mathtt{g})$ are the energy-momentum$1$-form fields) is represented by a
Lorentzian spacetime $\mathfrak{M}=(M,%
%TCIMACRO{\TeXButton{slg}{\slg}}%
%BeginExpansion
\slg
%EndExpansion
,D,\tau_{g},\uparrow)$. Let $\mathfrak{e\ }$be a constant with the physical
dimension of an electromagnetic potential and let
\begin{equation}
A=\mathfrak{e}K \label{potential}%
\end{equation}
be an electromagnetic potential.

Suppose now that a probe electromagnetic field $F=dA\in\sec\bigwedge
^{2}T^{\ast}M\hookrightarrow\mathcal{C\ell(}M,\mathtt{g})$ generated by a
(probe) current $J_{e}\in\sec\bigwedge^{1}T^{\ast}M\hookrightarrow
\mathcal{C\ell(}M,\mathtt{g})$ , lives and develops its dynamics in
$\mathfrak{M}$. Then we assume as usual that $F$ satisfies Maxwell equations
\begin{equation}
dF=0,\text{ }\delta F=-J_{e} \label{t15}%
\end{equation}
which taking into account the definition of the Dirac operator (Eq.(\ref{12}))
can be written as a single equation
\begin{equation}
{%
%TCIMACRO{\TeXButton{dirac}{\mbox{\boldmath$\partial$}}}%
%BeginExpansion
\mbox{\boldmath$\partial$}%
%EndExpansion
}F=J_{e}. \label{t16}%
\end{equation}

Now, it is usual in electrodynamics problems to work \ with the potential $A$
and fix the Lorenz gauge $\delta A=0$. This is done, e.g., in the classical
Eddington book \cite{eddington}. Taking into account Eq.(\ref{t1}) of
Proposition 1 we thus have that in a spacetime $\mathfrak{M}=(M,%
%TCIMACRO{\TeXButton{slg}{\slg}}%
%BeginExpansion
\slg
%EndExpansion
,D,\tau_{g},\uparrow)$ a probe electromagnetic field \ $F=dA$ such that
$\mathbf{A}=\mathtt{g}(A,$ $)$ is a Killing vector field is such that $\delta
A=0$, i.e., it is in the Lorentz gauge. Moreover. using Eq.(\ref{t14}) we see
that\ the current $J_{e}=2\mathcal{R}_{\beta}A^{\beta}$ is of the
superconductor type, i.e., proportional to the potential $A$. At the spacetime
points where the Ricci tensor is zero we necessarily have a null current.

This result is important since permit us to determine for each Killing vector
in $\mathfrak{M}$ a special current \ of superconductor type.

These results shows that in GR the potential $A$ appears to acquire a status
that it does not have in Special Relativity. In the reamining of the paper we
study \ further consequences of supposing that $A$ is proportional to a
Killing vector field

\begin{remark}
We recall that if $L\in\sec\bigwedge^{1}T^{\ast}M\hookrightarrow
\mathcal{C\ell(}M,\mathtt{g})$ is such that $\delta L=0$ we have that%
\begin{equation}
\frac{1}{2}g^{\kappa\mu}\left(  D_{\kappa}L_{\mu}+D_{\mu}L_{k}\right)  =0,
\label{t16b}%
\end{equation}
which however does \textit{not} implies in general that $D_{\kappa}L_{\mu
}+D_{\mu}L_{k}=0$, i.e., that $\mathbf{L}=\mathtt{g}(L,$ $)$ is a Killing
vector field.
\end{remark}

\section{The Spacetime $\mathfrak{M}$ Generated by an Ideal Current
Interacting with the Electromagnetic Field and the Gravitational Field}

In this section we analyze the dynamics of a coupled system consisting of an
ideal charged matter field plus electromagnetic field and the gravitational
field. For simplicity we restrict ourselves to the case of an incompressible
charged and frictionless fluid represented by a velocity field $\mathbf{V}%
\in\sec TM$ with $%
%TCIMACRO{\TeXButton{slg}{\slg}}%
%BeginExpansion
\slg
%EndExpansion
(\mathbf{V},\mathbf{V)}=1$ and such that each one of its integral lines, say
$\sigma:\tau\mapsto\sigma(\tau)\in M$ is such that $\left.  \mathbf{V}%
\right\vert _{\sigma}=\sigma_{\ast}=d/d\tau$ obeys the Lorentz force equation,
which writing $v=%
%TCIMACRO{\TeXButton{slg}{\slg}}%
%BeginExpansion
\slg
%EndExpansion
(\sigma_{\ast},$ $)$ reads:%
\begin{equation}
D_{\sigma_{\star}}v=\frac{e}{m}v\lrcorner F, \label{cs1}%
\end{equation}
where $e$ and $m$ are the charge and mass of the charged particles composing
the ideal charged fluid and $F$ is the total electromagnetic field generated
by it.

Introducing the velocity $1$-form field $V=%
%TCIMACRO{\TeXButton{slg}{\slg}}%
%BeginExpansion
\slg
%EndExpansion
(\mathbf{V},$ $)$ and using the noticeable identity given by Eq.(\ref{14A}) we
can write the first member of Eq.(\ref{cs1}) as%
\begin{align}
D_{\sigma_{\star}}v  &  =v^{\mu}D_{\partial_{\mu}}v=\left.  (V\lrcorner{%
%TCIMACRO{\TeXButton{dirac}{\mbox{\boldmath$\partial$}}}%
%BeginExpansion
\mbox{\boldmath$\partial$}%
%EndExpansion
)}V\right\vert _{\sigma}\label{cs2}\\
&  =\left.  V\lrcorner({%
%TCIMACRO{\TeXButton{dirac}{\mbox{\boldmath$\partial$}}}%
%BeginExpansion
\mbox{\boldmath$\partial$}%
%EndExpansion
\wedge}V{)}\right\vert _{\sigma}%
\end{align}
and thus since $F=dA$ and ${%
%TCIMACRO{\TeXButton{dirac}{\mbox{\boldmath$\partial$}}}%
%BeginExpansion
\mbox{\boldmath$\partial$}%
%EndExpansion
\wedge}V=dV$, Eq.(\ref{cs1}) implies that the velocity field satisfies the
equation
\begin{equation}
V\lrcorner\lbrack d(mV-eA){]=0.} \label{cs3}%
\end{equation}
A sufficient condition for the validity of Eq.(\ref{cs3}) is the existence of
a $0$-form field $S$ such that%
\[
mV-eA=dS.
\]
Then,%
\begin{equation}
dS+eA=mV \label{cs4}%
\end{equation}
and
\begin{equation}
(dS+eA)^{2}=m^{2} \label{cs5}%
\end{equation}
which we recognize as the classical Hamilton-Jacobi equation. Before
proceeding we recall that since $\delta V=0$ for a perfect incompressible
charged fluid, we get form Eq.(\ref{cs4})%
\begin{equation}
\delta dS+d\delta S+e\delta A=0, \label{cs4a}%
\end{equation}
i.e.,
\begin{equation}
{%
%TCIMACRO{\TeXButton{dirac}{\mbox{\boldmath$\partial$}}}%
%BeginExpansion
\mbox{\boldmath$\partial$}%
%EndExpansion
}^{2}S=e\delta A \label{cs4b}%
\end{equation}
\ 

\begin{remark}
Eq.\emph{(\ref{cs4b})} implies that the charged particle fluid action $S$ when
the potential $A$ \textit{is} in the Lorenz gauge satisfies a homogeneous wave
equation, even if $A$ is \textit{not} proportional to a Killing vector field.
\end{remark}

Then in what follows we call $S$ the action of the ideal charged fluid.
Moreover, since we are here more interested in the structure of the field
equations, we choose the mass and the charge of the fluid particles to be in
our system of units $m=1$ and $e=1$. Also the gravitational constant is $1$ in
our units.

Now, the Lagrangian density for the charged fluid model interacting with the
electromagnetic field and the gravitational field is given by
\begin{equation}
\mathcal{L=-}\frac{1}{2}J_{e}\mathbf{\wedge\star}J_{e}-\frac{1}{2}F\wedge\star
F+\mathcal{L}_{EH}, \label{cs6}%
\end{equation}
where%
\begin{equation}
\mathcal{L}_{EH}=\frac{1}{2}[\mathcal{R}_{\mu\nu}\wedge\star(%
%TCIMACRO{\TeXButton{bg}{\mbox{\boldmath{$\gamma$}}}}%
%BeginExpansion
\mbox{\boldmath{$\gamma$}}%
%EndExpansion
^{\mu}\wedge%
%TCIMACRO{\TeXButton{bg}{\mbox{\boldmath{$\gamma$}}}}%
%BeginExpansion
\mbox{\boldmath{$\gamma$}}%
%EndExpansion
^{\nu})] \label{CS7}%
\end{equation}
is the Einstein-Hilbert Lagrangian density and
\begin{equation}
J_{e}=(dS+A). \label{cs8}%
\end{equation}
The equations of motion resulting from the principle of stationary action are
\cite{thirring}:%
\begin{align}
\delta J_{e}  &  =0,\label{cs9}\\
\delta F  &  =-J_{e},\label{cs10}\\
\star\mathcal{G}_{\alpha}  &  =-\star T_{\alpha}-\star t_{\alpha}.
\label{cs11}%
\end{align}
where%
\begin{equation}
\star T_{\beta}=-\frac{1}{2}\star(F%
%TCIMACRO{\TeXButton{bg}{\mbox{\boldmath{$\gamma$}}}}%
%BeginExpansion
\mbox{\boldmath{$\gamma$}}%
%EndExpansion
_{\beta}F)
\end{equation}
are the energy-momentum $3$-form fields of the electromagnetic field (see,
e.g., \cite{rodoliv2007,notte} ) and the
\begin{align}
\star t_{\alpha}  &  =\frac{1}{2}[(%
%TCIMACRO{\TeXButton{bg}{\mbox{\boldmath{$\gamma$}}}}%
%BeginExpansion
\mbox{\boldmath{$\gamma$}}%
%EndExpansion
_{\alpha}\cdot J_{e})\wedge\star J_{e}+J_{e}\wedge(%
%TCIMACRO{\TeXButton{bg}{\mbox{\boldmath{$\gamma$}}}}%
%BeginExpansion
\mbox{\boldmath{$\gamma$}}%
%EndExpansion
_{\alpha}\cdot\star J_{e})\nonumber\\
&  =\frac{1}{2}\star(J_{e}%
%TCIMACRO{\TeXButton{bg}{\mbox{\boldmath{$\gamma$}}}}%
%BeginExpansion
\mbox{\boldmath{$\gamma$}}%
%EndExpansion
_{\alpha}J_{e}) \label{cs12}%
\end{align}
are the energy-momentum $3$-form fields of the ideal charged fluid. Note that
Eq. (\ref{cs9}) reproduces trivially Eq.(\ref{cs4}).

Now, it is easy to verify that Einstein equations implies that
\begin{equation}
R=J_{e}^{2}. \label{cse}%
\end{equation}
Then, we can rewrite Eq.(\ref{t12}) as%

\begin{align}
d\star dA  &  =-2\star A^{\beta}(\mathcal{R}_{\beta}-\frac{1}{2}R%
%TCIMACRO{\TeXButton{bg}{\mbox{\boldmath{$\gamma$}}}}%
%BeginExpansion
\mbox{\boldmath{$\gamma$}}%
%EndExpansion
_{\beta})-A^{\beta}R\star%
%TCIMACRO{\TeXButton{bg}{\mbox{\boldmath{$\gamma$}}}}%
%BeginExpansion
\mbox{\boldmath{$\gamma$}}%
%EndExpansion
_{\beta}\nonumber\\
&  =-\star2A^{\beta}\mathcal{G}_{\mathcal{\beta}}+J_{e}^{2}\star A\nonumber\\
&  =2\star\lbrack-\frac{1}{2}FAF+\frac{1}{2}J_{e}AJ_{e})+J_{e}^{2}\star
A\nonumber\\
&  =\star-(FAF)-\star AJ_{e}^{2}+2(A\cdot J_{e})\star J_{e}+J_{e}^{2}\star
A\nonumber\\
&  =\star-(FAF)+2(A\cdot J_{e})\star J_{e} \label{cs14}%
\end{align}
from where, since
\begin{equation}
\delta F=-J_{e},\text{ }dF=0 \label{ma}%
\end{equation}
we see that the current $J_{e}$ must satisfy the following equation
\begin{equation}
J_{e}=-FA\tilde{F}+2(A\cdot J_{e})J_{e}. \label{cs15}%
\end{equation}
If we make the exterior multiplication of both members of Eq.(\ref{cs15}) by
$J_{e}$ we get that $(FA\tilde{F})\wedge J_{e}=0$, from where taking into
account that $B=FA\tilde{F}\in\sec%
%TCIMACRO{\dbigwedge \nolimits^{1}}%
%BeginExpansion
{\displaystyle\bigwedge\nolimits^{1}}
%EndExpansion
T^{\star}M$ implies that
\begin{equation}
J_{e}=fFAF=-fB, \label{cs16}%
\end{equation}
where $f$ is a scalar function that in order for Eq.(\ref{cs15}) to be
satisfied must solve, when $B^{2}\neq0$, the quadratic equation%
\begin{equation}
2(A\cdot B)f^{2}+f+1=0, \label{cs17}%
\end{equation}
which has real roots only if $(A\cdot B)^{2}\leq\frac{1}{8}$. We then have the

\begin{proposition}
The Lorentzian spacetime $\mathfrak{M}=(M,%
%TCIMACRO{\TeXButton{slg}{\slg}}%
%BeginExpansion
\slg
%EndExpansion
,D,\tau_{g},\uparrow)$ where an incompressible charged fluid\ described by an
action $S$ generates an electromagnetic field \ $F=dA$ such that
$\mathbf{A}=\mathtt{g}(A,)$ is proportional to a Killing vector field and both
are in interaction with the gravitational field is such that and the current
$\ $is given by $J_{e}=fFAF$ where $f$ is a solution of the algebraic
quadratic equation given by \emph{Eq.(\ref{cs17}) with }$[A\cdot(FAF)]^{2}%
\leq\frac{1}{8}.$Moreover, Einstein equations are equivalent to Maxwell
equations given by \emph{Eq.(\ref{cs14})}.
\end{proposition}

\section{Conclusions}

In this paper using the Clifford bundle formalism and a proposition
(Proposition 1) of differential geometry which shows that if $\mathbf{K}$ is a
Killing vector field on a Lorentzian manifold $(M,%
%TCIMACRO{\TeXButton{slg}{\slg}}%
%BeginExpansion
\slg
%EndExpansion
)$ then the form field $K=$ $%
%TCIMACRO{\TeXButton{slg}{\slg}}%
%BeginExpansion
\slg
%EndExpansion
(\mathbf{K},$ $)$ satisfies $\delta K=0$ and a wave equation given by
Eq.(\ref{t2}) in terms of the covariant D' Alembertian applied to $K$ . We
also showed that the Ricci operator, which can be defined only in the Clifford
bundle of differential forms applied to $K$ is equal to the covariant
D'Alembertian \ applied to $K$. We analyzed morevoer the structure of Maxwell
equations in a Lorentzian spacetime when the potential is proportional to $K$,
$A=\mathfrak{e}K$ and thus satisfies the Lorenz gauge $\delta A=0$. The
explicit form of the current which generates the electromagnetic field has
been calculated and resulted proportional to $FAF$. Next we studied the
\textit{structure} of the spacetime generated by the interaction of the a
perfect charged fluid described by action $S$, its electromagnetic field
$F=dA$, with $A=\mathfrak{e}K$ and the gravitational field . We found that
Einstein equations for this case is represented by Maxwell equations with a
current given by Proposition 4, i.e., $J_{e}=f$ $FAF$, where $f$ is a scalar
function which is solution of a \ well determined quadratic equation.

\appendix

\section{Clifford Bundle Formalism}

Let $\mathfrak{M}=(M,%
%TCIMACRO{\TeXButton{slg}{\slg}}%
%BeginExpansion
\slg
%EndExpansion
,D,\tau_{g},\uparrow)$ be an arbitrary Lorentzian spacetime.\ The quadruple
$(M,%
%TCIMACRO{\TeXButton{slg}{\slg}}%
%BeginExpansion
\slg
%EndExpansion
,\tau_{g},\uparrow)$ denotes a four-dimensional time-oriented and
space-oriented Lorentzian manifold \cite{rodoliv2007,sawu}. This means that $%
%TCIMACRO{\TeXButton{slg}{\slg}}%
%BeginExpansion
\slg
%EndExpansion
\in\sec T_{2}^{0}M$ is a Lorentzian metric of signature (1,3), $\tau_{g}%
\in\sec\bigwedge{}^{4}T^{\ast}M$ and $\uparrow$ is a time-orientation (see
details, e.g., in \cite{sawu}). Here, $T^{\ast}M$ [$TM$] is the cotangent
[tangent] bundle. $T^{\ast}M=\cup_{x\in M}T_{x}^{\ast}M$, $TM=\cup_{x\in
M}T_{x}M$, and $T_{x}M\simeq T_{x}^{\ast}M\simeq\mathbb{R}^{1,3}$, where
$\mathbb{R}^{1,3}$ is the Minkowski vector space\footnote{Not to be confused
with Minkowski spacetime \cite{sawu}.}. $D$ is the Levi-Civita connection of $%
%TCIMACRO{\TeXButton{slg}{\slg}}%
%BeginExpansion
\slg
%EndExpansion
$, i.e., it is \ metric compatible connection, i.e.\/, $D%
%TCIMACRO{\TeXButton{slg}{\slg}}%
%BeginExpansion
\slg
%EndExpansion
=0$, and in general, $\mathbf{R}=\mathbf{R}^{D}\neq0$, and $\Theta=\Theta
^{D}=0$, $\mathbf{R}$ and $\Theta$ being respectively the curvature and
torsion tensors of the connection. Minkowski spacetime is the particular case
of a Lorentzian spacetime for which $\mathbf{R}=0$, $\Theta=0$, and
$M\simeq\mathbb{R}^{4}$. Let $\mathtt{g}\in\sec T_{0}^{2}M$ be the metric of
the \textit{cotangent bundle}. The Clifford bundle of differential forms
$\mathcal{C}\!\ell(M,\mathtt{g})$ is the bundle of algebras, i.e.,
$\mathcal{C}\ell(M,\mathtt{g})=\cup_{x\in M}\mathcal{C}\!\ell(T_{x}^{\ast
}M,\mathtt{g})$, where $\forall x\in M$, $\mathcal{C}\!\ell(T_{x}^{\ast
}M,\mathtt{g})=\mathbb{R}_{1,3}$, the so called \emph{spacetime} \emph{algebra
}\cite{rodoliv2007}. Recall also that $\mathcal{C}\ell(M,\mathtt{g})$ is a
vector bundle associated to the \emph{orthonormal frame bundle}, i.e.,
$\mathcal{C}\ell(M,\mathtt{g})$ $=P_{\mathrm{SO}_{(1,3)}^{e}}(M)\times
_{\mathrm{Ad}}\mathcal{C}l_{1,3}$ \cite{lawmi,moro}. For any $x\in M$,
$\mathcal{C}\ell(T_{x}^{\ast}M,\left.  \mathtt{g}\right\vert _{x})$ as a
linear space over the real field $\mathbb{R}$ is isomorphic to the Cartan
algebra $\bigwedge T_{x}^{\ast}M$ of the cotangent space. $\bigwedge
T_{x}^{\ast}M=\oplus_{k=0}^{4}\bigwedge^{k}T_{x}^{\ast}M$, where
$\bigwedge^{k}T_{x}^{\ast}M$ is the $\binom{4}{k}$-dimensional space of
$k$-forms. Then, sections of $\mathcal{C}\ell(M,\mathtt{g})$ can be
represented as a sum of non homogeneous differential forms, that will be
called Clifford (multiform) fields. In the Clifford bundle formalism, of
course, arbitrary basis can be used, but in this short review of the main
ideas of the Clifford calculus we use orthonormal basis. Let then
$\{\mathbf{e}_{\mathbf{a}}\}$ be an orthonormal basis for $TU\subset TM$,
i.e., $\mathtt{g}(\mathbf{e}_{\mathbf{a}},\mathbf{e}_{\mathbf{a}}%
)=\eta_{\mathbf{ab}}=\mathrm{diag}(1,-1,-1,-1)$. Let $%
%TCIMACRO{\TeXButton{bt}{\mbox{\boldmath{$\theta$}}}}%
%BeginExpansion
\mbox{\boldmath{$\theta$}}%
%EndExpansion
^{\mathbf{a}}\in\sec\bigwedge^{1}T^{\ast}M\hookrightarrow\sec\mathcal{C}%
\ell(M,\mathtt{g})$ ($\mathbf{a}=0,1,2,3$) be such that the set $\{%
%TCIMACRO{\TeXButton{bt}{\mbox{\boldmath{$\theta$}}}}%
%BeginExpansion
\mbox{\boldmath{$\theta$}}%
%EndExpansion
^{\mathbf{a}}\}$ is the dual basis of $\{\mathbf{e}_{\mathbf{a}}\}$.

\subsection{Clifford Product}

The fundamental \emph{Clifford product} (in what follows to be denoted by
juxtaposition of symbols) is generated by
\begin{equation}%
%TCIMACRO{\TeXButton{bt}{\mbox{\boldmath{$\theta$}}}}%
%BeginExpansion
\mbox{\boldmath{$\theta$}}%
%EndExpansion
^{\mathbf{a}}%
%TCIMACRO{\TeXButton{bt}{\mbox{\boldmath{$\theta$}}}}%
%BeginExpansion
\mbox{\boldmath{$\theta$}}%
%EndExpansion
^{\mathbf{b}}+%
%TCIMACRO{\TeXButton{bt}{\mbox{\boldmath{$\theta$}}}}%
%BeginExpansion
\mbox{\boldmath{$\theta$}}%
%EndExpansion
^{\mathbf{b}}%
%TCIMACRO{\TeXButton{bt}{\mbox{\boldmath{$\theta$}}}}%
%BeginExpansion
\mbox{\boldmath{$\theta$}}%
%EndExpansion
^{\mathbf{a}}=2\eta^{\mathbf{ab}} \label{cl}%
\end{equation}
and if $\mathcal{C}\in\sec\mathcal{C}\ell(M,\mathtt{g})$ we have%

\begin{equation}
\mathcal{C}=s+v_{\mathbf{a}}%
%TCIMACRO{\TeXButton{bt}{\mbox{\boldmath{$\theta$}}}}%
%BeginExpansion
\mbox{\boldmath{$\theta$}}%
%EndExpansion
^{\mathbf{a}}+\frac{1}{2!}f_{\mathbf{ab}}%
%TCIMACRO{\TeXButton{bt}{\mbox{\boldmath{$\theta$}}}}%
%BeginExpansion
\mbox{\boldmath{$\theta$}}%
%EndExpansion
^{\mathbf{a}}%
%TCIMACRO{\TeXButton{bt}{\mbox{\boldmath{$\theta$}}}}%
%BeginExpansion
\mbox{\boldmath{$\theta$}}%
%EndExpansion
^{\mathbf{b}}+\frac{1}{3!}t_{\mathbf{abc}}%
%TCIMACRO{\TeXButton{bt}{\mbox{\boldmath{$\theta$}}}}%
%BeginExpansion
\mbox{\boldmath{$\theta$}}%
%EndExpansion
^{\mathbf{a}}%
%TCIMACRO{\TeXButton{bt}{\mbox{\boldmath{$\theta$}}}}%
%BeginExpansion
\mbox{\boldmath{$\theta$}}%
%EndExpansion
^{\mathbf{b}}%
%TCIMACRO{\TeXButton{bt}{\mbox{\boldmath{$\theta$}}}}%
%BeginExpansion
\mbox{\boldmath{$\theta$}}%
%EndExpansion
^{\mathbf{c}}+p%
%TCIMACRO{\TeXButton{bt}{\mbox{\boldmath{$\theta$}}}}%
%BeginExpansion
\mbox{\boldmath{$\theta$}}%
%EndExpansion
^{5}\;, \label{3}%
\end{equation}
where $\tau_{g}=%
%TCIMACRO{\TeXButton{bt}{\mbox{\boldmath{$\theta$}}}}%
%BeginExpansion
\mbox{\boldmath{$\theta$}}%
%EndExpansion
^{5}=%
%TCIMACRO{\TeXButton{bt}{\mbox{\boldmath{$\theta$}}}}%
%BeginExpansion
\mbox{\boldmath{$\theta$}}%
%EndExpansion
^{0}%
%TCIMACRO{\TeXButton{bt}{\mbox{\boldmath{$\theta$}}}}%
%BeginExpansion
\mbox{\boldmath{$\theta$}}%
%EndExpansion
^{\mathbf{1}}%
%TCIMACRO{\TeXButton{bt}{\mbox{\boldmath{$\theta$}}}}%
%BeginExpansion
\mbox{\boldmath{$\theta$}}%
%EndExpansion
^{\mathbf{2}}%
%TCIMACRO{\TeXButton{bt}{\mbox{\boldmath{$\theta$}}}}%
%BeginExpansion
\mbox{\boldmath{$\theta$}}%
%EndExpansion
^{\mathbf{3}}$ is the volume element and $s$, $v_{\mathbf{a}}$,
$f_{\mathbf{ab}}$, $t_{\mathbf{abc}}$, $p\in\sec\bigwedge^{0}T^{\ast
}M\hookrightarrow\sec\mathcal{C}\!\ell(M,\mathtt{g})$.

For $A_{r}\in\sec\bigwedge^{r}T^{\ast}M\hookrightarrow\sec\mathcal{C}%
\!\ell(M,\mathtt{g}),B_{s}\in\sec\bigwedge^{s}T^{\ast}M\hookrightarrow
\sec\mathcal{C}\!\ell(M,\mathtt{g})$ we define the \emph{exterior product} in
$\mathcal{C}\!\ell(M,\mathtt{g})$ \ ($\forall r,s=0,1,2,3)$ by
\begin{equation}
A_{r}\wedge B_{s}=\langle A_{r}B_{s}\rangle_{r+s}, \label{5}%
\end{equation}
where $\langle\;\;\rangle_{k}$ is the component in $\bigwedge^{k}T^{\ast}M$ of
the Clifford field. Of course, $A_{r}\wedge B_{s}=(-1)^{rs}B_{s}\wedge A_{r}$,
and the exterior product is extended by linearity to all sections of
$\mathcal{C}\!\ell(M,\mathtt{g})$.

Let $A_{r}\in\sec\bigwedge^{r}T^{\ast}M\hookrightarrow\sec\mathcal{C}%
\!\ell(M,\mathtt{g}),B_{s}\in\sec\bigwedge^{s}T^{\ast}M\hookrightarrow
\sec\mathcal{C}\!\ell(M,\mathtt{g})$. We define a \emph{scalar product
}in\emph{\ }$\mathcal{C}\!\ell(M,\mathtt{g})$ (denoted by $\cdot$) as follows:

(i) For $a,b\in\sec\bigwedge^{1}T^{\ast}M\hookrightarrow\sec\mathcal{C}%
\!\ell(M,\mathtt{g}),$%
\begin{equation}
a\cdot b=\frac{1}{2}(ab+ba)=\mathtt{g}(a,b). \label{4}%
\end{equation}

(ii) For $A_{r}=a_{1}\wedge...\wedge a_{r},B_{r}=b_{1}\wedge...\wedge b_{r}$,
$a_{i},b_{j}\in\sec\bigwedge^{1}T^{\ast}M\hookrightarrow\sec\mathcal{C}%
\!\ell(M,\mathtt{g})$, $i,j=1,...,r,$
\begin{align}
A_{r}\cdot B_{r}  &  =(a_{1}\wedge...\wedge a_{r})\cdot(b_{1}\wedge...\wedge
b_{r})\nonumber\\
&  =\left\vert
\begin{array}
[c]{lll}%
a_{1}\cdot b_{1} & .... & a_{1}\cdot b_{r}\\
.......... & .... & ..........\\
a_{r}\cdot b_{1} & .... & a_{r}\cdot b_{r}%
\end{array}
\right\vert . \label{6}%
\end{align}

We agree that if $r=s=0$, the scalar product is simply the ordinary product in
the real field.

Also, if $r\neq s$, then $A_{r}\cdot B_{s}=0$. Finally, the scalar product is
extended by linearity for all sections of $\mathcal{C}\!\ell(M,\mathtt{g})$.

For $r\leq s$, $A_{r}=a_{1}\wedge...\wedge a_{r}$, $B_{s}=b_{1}\wedge...\wedge
b_{s\text{ }}$, we define the \textit{left contraction} $\lrcorner
:(A_{r},B_{s})\mapsto A_{r}\lrcorner B_{s}$ by
\begin{equation}
A_{r}\lrcorner B_{s}=%
%TCIMACRO{\dsum \limits_{i_{1}\,<...\,<i_{r}}}%
%BeginExpansion
{\displaystyle\sum\limits_{i_{1}\,<...\,<i_{r}}}
%EndExpansion
\epsilon^{i_{1}...i_{s}}(a_{1}\wedge...\wedge a_{r})\cdot(b_{_{i_{1}}}%
\wedge...\wedge b_{i_{r}})^{\sim}b_{i_{r}+1}\wedge...\wedge b_{i_{s}}
\label{7}%
\end{equation}
where $\sim$ is the reverse mapping (\emph{reversion}) defined by
$\symbol{126}:\sec\mathcal{C}\!\ell(M,\mathtt{g})\rightarrow\sec
\mathcal{C}\!\ell(M,\mathtt{g})$. For any $X=%
%TCIMACRO{\dbigoplus \nolimits_{p=0}^{4}}%
%BeginExpansion
{\displaystyle\bigoplus\nolimits_{p=0}^{4}}
%EndExpansion
X_{p}$,$X_{p}\in\sec%
%TCIMACRO{\dbigwedge \nolimits^{p}}%
%BeginExpansion
{\displaystyle\bigwedge\nolimits^{p}}
%EndExpansion
T^{\ast}M\hookrightarrow\sec\mathcal{C}\!\ell(M,\mathtt{g})$,
\begin{equation}
\tilde{X}=%
%TCIMACRO{\dsum \limits_{p=0}^{4}}%
%BeginExpansion
{\displaystyle\sum\limits_{p=0}^{4}}
%EndExpansion
\text{ }\tilde{X}_{p}=%
%TCIMACRO{\dsum \limits_{p=0}^{4}}%
%BeginExpansion
{\displaystyle\sum\limits_{p=0}^{4}}
%EndExpansion
(-1)^{\frac{1}{2}k(k-1)}X_{p}.
\end{equation}
We agree that for $\alpha,\beta\in\sec\bigwedge^{0}T^{\ast}M$ the contraction
is the ordinary (pointwise) product in the real field and that if $\alpha
\in\sec\bigwedge^{0}T^{\ast}M$, $X_{r}\in\sec\bigwedge^{r}T^{\ast}M,Y_{s}%
\in\sec\bigwedge^{s}T^{\ast}M\hookrightarrow\sec\mathcal{C}\!\ell
(M,\mathtt{g})$ then $(\alpha X_{r})\lrcorner B_{s}=X_{r}\lrcorner(\alpha
Y_{s})$. Left contraction is extended by linearity to all pairs of sections of
$\mathcal{C}\!\ell(M,\mathtt{g})$, i.e., for $X,Y\in\sec\mathcal{C}%
\!\ell(M,\mathtt{g})$%
\begin{equation}
X\lrcorner Y=\sum_{r,s}\langle X\rangle_{r}\lrcorner\langle Y\rangle_{s},\quad
r\leq s. \label{9}%
\end{equation}

It is also necessary to introduce the operator of \emph{right contraction}
denoted by $\llcorner$. The definition is obtained from the one presenting the
left contraction with the imposition that $r\geq s$ and taking into account
that now if $A_{r}\in\sec\bigwedge^{r}T^{\ast}M,$ $B_{s}\in\sec\bigwedge
^{s}T^{\ast}M$ then $A_{r}\llcorner(\alpha B_{s})=(\alpha A_{r})\llcorner
B_{s}$. See also the third formula in Eq.(\ref{10}).

The main formulas used in this paper can be obtained from the following ones
\begin{align}
a\mathcal{B}_{s}  &  =a\lrcorner\mathcal{B}_{s}+a\wedge\mathcal{B}%
_{s},\;\;\mathcal{B}_{s}a=\mathcal{B}_{s}\llcorner a+\mathcal{B}_{s}\wedge
a,\nonumber\\
a\lrcorner\mathcal{B}_{s}  &  =\frac{1}{2}(a\mathcal{B}_{s}-(-1)^{s}%
\mathcal{B}_{s}a),\nonumber\\
\mathcal{A}_{r}\lrcorner\mathcal{B}_{s}  &  =(-1)^{r(s-r)}\mathcal{B}%
_{s}\llcorner\mathcal{A}_{r},\nonumber\\
a\wedge\mathcal{B}_{s}  &  =\frac{1}{2}(a\mathcal{B}_{s}+(-1)^{s}%
\mathcal{B}_{s}a),\nonumber\\
\mathcal{A}_{r}\mathcal{B}_{s}  &  =\langle\mathcal{A}_{r}\mathcal{B}%
_{s}\rangle_{|r-s|}+\langle\mathcal{A}_{r}\mathcal{B}_{s}\rangle
_{|r-s|+2}+...+\langle\mathcal{A}_{r}\mathcal{B}_{s}\rangle_{|r+s|}\nonumber\\
&  =\sum\limits_{k=0}^{m}\langle\mathcal{A}_{r}\mathcal{B}_{s}\rangle
_{|r-s|+2k}\text{ }\nonumber\\
\mathcal{A}_{r}\cdot\mathcal{B}_{r}  &  =\mathcal{B}_{r}\cdot\mathcal{A}%
_{r}=\widetilde{\mathcal{A}}_{r}\text{ }\lrcorner\mathcal{B}_{r}%
=\mathcal{A}_{r}\llcorner\widetilde{\mathcal{B}}_{r}=\langle\widetilde
{\mathcal{A}}_{r}\mathcal{B}_{r}\rangle_{0}=\langle\mathcal{A}_{r}%
\widetilde{\mathcal{B}}_{r}\rangle_{0}. \label{10}%
\end{align}
Two other important identities used in the main text are:%

\begin{align}
a\lrcorner(\mathcal{X}\wedge\mathcal{Y})  &  =(a\lrcorner\mathcal{X}%
)\wedge\mathcal{Y}+\mathcal{\hat{X}}\wedge(a\lrcorner\mathcal{Y}%
),\label{T54}\\
A\lrcorner(B\lrcorner C)  &  =(A\wedge B)\lrcorner C, \label{T55}%
\end{align}
for any $a\in\sec%
%TCIMACRO{\dbigwedge \nolimits^{1}}%
%BeginExpansion
{\displaystyle\bigwedge\nolimits^{1}}
%EndExpansion
T^{\ast}M\hookrightarrow\mathcal{C}\ell(M,\mathtt{g})$ and $\mathcal{X}%
,\mathcal{Y}\in\sec%
%TCIMACRO{\dbigwedge }%
%BeginExpansion
{\displaystyle\bigwedge}
%EndExpansion
T^{\ast}M\hookrightarrow\mathcal{C}\ell(M,\mathtt{g})$, and for any
$A,B,C\in\sec\bigwedge T^{\ast}M\hookrightarrow\mathcal{C}\ell(M,\mathtt{g})$.

\subsubsection{Hodge Star Operator}

Let $\star$ be the Hodge star operator, i.e., the mapping $\star:%
%TCIMACRO{\dbigwedge \nolimits^{k}}%
%BeginExpansion
{\displaystyle\bigwedge\nolimits^{k}}
%EndExpansion
T^{\ast}M\rightarrow%
%TCIMACRO{\dbigwedge \nolimits^{4-k}}%
%BeginExpansion
{\displaystyle\bigwedge\nolimits^{4-k}}
%EndExpansion
T^{\ast}M,$ $A_{k}\mapsto\star A_{k}$. For $A_{k}\in\sec\bigwedge^{k}T^{\ast
}M\hookrightarrow\sec\mathcal{C}\!\ell(M,\mathtt{g})$ we have
\begin{equation}
\lbrack B_{k}\cdot A_{k}]\tau_{\mathtt{g}}=B_{k}\wedge\star A_{k},\forall
B_{k}\in\sec\bigwedge\nolimits^{k}T^{\ast}M\hookrightarrow\sec\mathcal{C}%
\!\ell(M,\mathtt{g}). \label{11a}%
\end{equation}
where $\tau_{\mathtt{g}}=\theta^{\mathbf{5}}\in\sec\bigwedge^{4}T^{\ast
}M\hookrightarrow\sec\mathcal{C}\!\ell(M,\mathtt{g})$ is a \emph{standard}
volume element. We have,
\begin{equation}
\star A_{k}=\widetilde{A}_{k}\tau_{\mathtt{g}}=\widetilde{A}_{k}\lrcorner
\tau_{\mathtt{g}}. \label{11b}%
\end{equation}
where as noted before, in this paper $\widetilde{\mathcal{A}}_{k}$ denotes the
\textit{reverse} of $\mathcal{A}_{k}$. Eq.(\ref{11b}) permits calculation of
Hodge duals very easily in an orthonormal basis for which $\tau_{\mathtt{g}}=%
%TCIMACRO{\TeXButton{bt}{\mbox{\boldmath{$\theta$}}}}%
%BeginExpansion
\mbox{\boldmath{$\theta$}}%
%EndExpansion
^{\mathbf{5}}$. Let $\{\vartheta^{\alpha}\}$ be the dual basis of
$\{e_{\alpha}\}$ (i.e., it is a basis for $T^{\ast}U\equiv\bigwedge
\nolimits^{1}T^{\ast}U$) which is either \textit{orthonormal} or a
\textit{coordinate basis}. Then writing \texttt{g}$(\vartheta^{\alpha
},\vartheta^{\beta})=g^{\alpha\beta}$, with $g^{\alpha\beta}g_{\alpha\rho
}=\delta_{\rho}^{\beta}$, and $\vartheta^{\mu_{1}...\mu_{p}}=\vartheta
^{\mu_{1}}\wedge...\wedge\vartheta^{\mu_{p}}$, $\vartheta^{\nu_{p+1}...\nu
_{n}}=\vartheta^{\nu_{p+1}}\wedge...\wedge\vartheta^{\nu_{n}}$ we have from
Eq.(\ref{11b})
\begin{equation}
{}\star\vartheta^{\mu_{1}...\mu_{p}}=\frac{1}{(n-p)!}\sqrt{\left\vert
\mathbf{g}\right\vert }g^{\mu_{1}\nu_{1}}...g^{\mu_{p}\nu_{p}}\epsilon
_{\nu_{1}...\nu_{n}}\vartheta^{\nu_{p+1}...\nu_{n}}. \label{hodge dual}%
\end{equation}
where $\mathbf{g}$ denotes the determinant of the matrix with entries
$g_{\alpha\beta}=$\texttt{ }$%
%TCIMACRO{\TeXButton{slg}{\slg}}%
%BeginExpansion
\slg
%EndExpansion
(e_{\alpha},e_{\beta})$, i.e.,$\mathbf{g}=\det[g_{\alpha\beta}].$ We also
define the inverse $\star^{-1}$ of the Hodge dual operator, such that
\ $\star^{-1}\star=\star\star^{-1}=1$. It is given by:
\begin{align}
\star^{-1}  &  :\sec%
%TCIMACRO{\dbigwedge \nolimits^{n-r}}%
%BeginExpansion
{\displaystyle\bigwedge\nolimits^{n-r}}
%EndExpansion
T^{\ast}M\rightarrow\sec%
%TCIMACRO{\dbigwedge \nolimits^{r}}%
%BeginExpansion
{\displaystyle\bigwedge\nolimits^{r}}
%EndExpansion
T^{\ast}M,\nonumber\\
\star^{-1}  &  =(-1)^{r(n-r)}\mathrm{sgn}\text{ }\mathbf{g\,}\star, \label{h1}%
\end{align}
where \textrm{sgn }$\mathbf{g}=\mathbf{g}/|\mathbf{g}|$ denotes the sign of
the determinant $\mathbf{g}$.

Some useful identities (used in the text) involving the Hodge star operator,
the exterior product and contractions are:%

\begin{equation}%
\begin{array}
[c]{l}%
A_{r}\wedge\star B_{s}=B_{s}\wedge\star A_{r};\quad r=s\\
A_{r}\cdot\star B_{s}=B_{s}\cdot\star A_{r};\quad r+s=n\\
A_{r}\wedge\star B_{s}=(-1)^{r(s-1)}\star(\tilde{A}_{r}\lrcorner B_{s});\quad
r\leq s\\
A_{r}\lrcorner\star B_{s}=(-1)^{rs}\star(\tilde{A}_{r}\wedge B_{s});\quad
r+s\leq n\\
\star\tau_{g}=\mathrm{sign}\text{ }\mathbf{g};\quad\star1=\tau_{g}.
\end{array}
\label{440new}%
\end{equation}

\subsubsection{Dirac Operator Associated to a Levi-Civita Connection}

Let $d$ and $\delta$ be respectively the differential and Hodge codifferential
operators acting on sections of $\mathcal{C}\!\ell(M,\mathtt{g})$. If
$A_{p}\in\sec\bigwedge^{p}T^{\ast}M\hookrightarrow\sec\mathcal{C}%
\!\ell(M,\mathtt{g})$, then $\delta A_{p}=(-1)^{p}\star^{-1}d\star A_{p}$.

The Dirac operator acting on sections of $\mathcal{C}\!\ell(M,\mathtt{g})$
associated with the metric compatible connection $D$ is the invariant first
order differential operator
\begin{equation}
{\mbox{\boldmath$\partial$}}=\vartheta^{\mathbf{\alpha}}D_{e_{\alpha}},
\label{12}%
\end{equation}
where $\{e_{\mathbf{\alpha}}\}$ is an arbitrary (coordinate or
orthonormal)\emph{ basis} for $TU\subset TM$ and $\{\vartheta^{\mathbf{\alpha
}}\}$ is a basis for $T^{\ast}U\subset T^{\ast}M$ dual to the basis
$\{e_{\mathbf{\alpha}}\}$, i.e., $\vartheta^{\beta}(e_{\mathbf{\alpha}%
})=\delta_{\mathbf{\beta}}^{\mathbf{\alpha}}$, $\mathbf{\alpha}\mathbf{,\beta
}=0,1,2,3$. The reciprocal basis of $\{\vartheta^{\mathbf{\alpha}}\}$ is
denoted $\{\vartheta_{\mathbf{\alpha}}\}$ and we have $\vartheta
_{\mathbf{\alpha}}\cdot\vartheta_{\beta}=g_{\mathbf{\alpha\beta}}$. Also,
\begin{equation}
D_{e_{\mathbf{\alpha}}}\vartheta^{\mathbf{\beta}}=-L_{\alpha\mathbf{\lambda}%
}^{\mathbf{\beta}}\vartheta^{\lambda} \label{12n}%
\end{equation}
and we define the connection $1$-forms in the gauge defined by $\{\vartheta
^{\mathbf{\alpha}}\}$ as%
\begin{equation}
L_{\mathbf{\beta}}^{\mathbf{\alpha}}:=L_{\mathbf{\lambda\beta}}%
^{\mathbf{\alpha}}\vartheta^{\mathbf{\lambda}}. \label{12na}%
\end{equation}
We recall also that for an orthonormal basis it is usual to write
($\mathbf{a,b,c}=0,1,2,3$)%

\begin{equation}
D_{\mathbf{e}_{\mathbf{a}}}%
%TCIMACRO{\TeXButton{bt}{\mbox{\boldmath{$\theta$}}}}%
%BeginExpansion
\mbox{\boldmath{$\theta$}}%
%EndExpansion
^{\mathbf{b}}=-\omega_{\mathbf{ac}}^{\mathbf{b}}%
%TCIMACRO{\TeXButton{bt}{\mbox{\boldmath{$\theta$}}}}%
%BeginExpansion
\mbox{\boldmath{$\theta$}}%
%EndExpansion
^{\mathbf{c}},\text{ }\omega_{\mathbf{b}}^{\mathbf{a}}:=\omega_{\mathbf{cb}%
}^{\mathbf{a}}%
%TCIMACRO{\TeXButton{bt}{\mbox{\boldmath{$\theta$}}}}%
%BeginExpansion
\mbox{\boldmath{$\theta$}}%
%EndExpansion
^{\mathbf{c}}. \label{12nb}%
\end{equation}
Moreover, we write for an arbitrary tensor field $Y=Y_{\nu_{1}...\nu_{s}}%
^{\mu_{1}...\mu_{r}}%
%TCIMACRO{\TeXButton{bg}{\mbox{\boldmath{$\gamma$}}}}%
%BeginExpansion
\mbox{\boldmath{$\gamma$}}%
%EndExpansion
^{\nu_{1}}\otimes...\otimes%
%TCIMACRO{\TeXButton{bg}{\mbox{\boldmath{$\gamma$}}}}%
%BeginExpansion
\mbox{\boldmath{$\gamma$}}%
%EndExpansion
^{\nu_{s}}\otimes\partial_{\mu_{1}}\otimes...\otimes\partial_{\mu_{r}}$ in a
coordinate basis (and use the notation of Section 2 for the basis of the
tangent and cotangent bundles) $,$%
\begin{equation}
D_{\mathbf{e}_{\mathbf{\alpha}}}Y:=(D_{\alpha}Y_{\nu_{1}...\nu_{s}}^{\mu
_{1}...\mu_{r}})%
%TCIMACRO{\TeXButton{bg}{\mbox{\boldmath{$\gamma$}}}}%
%BeginExpansion
\mbox{\boldmath{$\gamma$}}%
%EndExpansion
^{\nu_{1}}\otimes...\otimes%
%TCIMACRO{\TeXButton{bg}{\mbox{\boldmath{$\gamma$}}}}%
%BeginExpansion
\mbox{\boldmath{$\gamma$}}%
%EndExpansion
^{\nu_{s}}\otimes\partial_{\mu_{1}}\otimes...\otimes\partial_{\mu_{r}}
\label{cd}%
\end{equation}

We have also the important results (see, e.g., \cite{rodoliv2007}) for the
Dirac operator associated with the Levi-Civita connection $D$ acting on the
sections of the Clifford bundle%

\begin{align}
{%
%TCIMACRO{\TeXButton{dirac}{\mbox{\boldmath$\partial$}}}%
%BeginExpansion
\mbox{\boldmath$\partial$}%
%EndExpansion
}A_{p}  &  ={%
%TCIMACRO{\TeXButton{dirac}{\mbox{\boldmath$\partial$}}}%
%BeginExpansion
\mbox{\boldmath$\partial$}%
%EndExpansion
}\wedge A_{p\,}+\,{%
%TCIMACRO{\TeXButton{dirac}{\mbox{\boldmath$\partial$}}}%
%BeginExpansion
\mbox{\boldmath$\partial$}%
%EndExpansion
}\lrcorner A_{p}=dA_{p}-\delta A_{p},\nonumber\\
{%
%TCIMACRO{\TeXButton{dirac}{\mbox{\boldmath$\partial$}}}%
%BeginExpansion
\mbox{\boldmath$\partial$}%
%EndExpansion
}\wedge A_{p}  &  =dA_{p},\hspace{0.1in}\,{%
%TCIMACRO{\TeXButton{dirac}{\mbox{\boldmath$\partial$}}}%
%BeginExpansion
\mbox{\boldmath$\partial$}%
%EndExpansion
}\lrcorner A_{p}=-\delta A_{p}. \label{13}%
\end{align}
We shall need the following identity valid for any $A,B\in\sec\bigwedge
^{1}T^{\ast}M\hookrightarrow\mathcal{C}\ell(M,\mathtt{g}),$%
\begin{equation}
{%
%TCIMACRO{\TeXButton{dirac}{\mbox{\boldmath$\partial$}}}%
%BeginExpansion
\mbox{\boldmath$\partial$}%
%EndExpansion
(}A\cdot B)={(A\cdot%
%TCIMACRO{\TeXButton{dirac}{\mbox{\boldmath$\partial$}}}%
%BeginExpansion
\mbox{\boldmath$\partial$}%
%EndExpansion
})B+(B\cdot{%
%TCIMACRO{\TeXButton{dirac}{\mbox{\boldmath$\partial$}}}%
%BeginExpansion
\mbox{\boldmath$\partial$}%
%EndExpansion
)}A+A\lrcorner({%
%TCIMACRO{\TeXButton{dirac}{\mbox{\boldmath$\partial$}}}%
%BeginExpansion
\mbox{\boldmath$\partial$}%
%EndExpansion
\wedge}B)+B\lrcorner({%
%TCIMACRO{\TeXButton{dirac}{\mbox{\boldmath$\partial$}}}%
%BeginExpansion
\mbox{\boldmath$\partial$}%
%EndExpansion
\wedge}A). \label{14A}%
\end{equation}

\subsection{Covariant D' Alembertian, Hodge D'Alembertian and Ricci Operators}

The square of the Dirac operator $\Diamond={%
%TCIMACRO{\TeXButton{dirac}{\mbox{\boldmath$\partial$}}}%
%BeginExpansion
\mbox{\boldmath$\partial$}%
%EndExpansion
}^{2}$ is called Hodge D'Alembertian and we have the following noticeable
formulas:%
\begin{equation}
{%
%TCIMACRO{\TeXButton{dirac}{\mbox{\boldmath$\partial$}}}%
%BeginExpansion
\mbox{\boldmath$\partial$}%
%EndExpansion
}^{2}=-d\delta-\delta d,\label{13bis}%
\end{equation}
and
\begin{equation}
{%
%TCIMACRO{\TeXButton{dirac}{\mbox{\boldmath$\partial$}}}%
%BeginExpansion
\mbox{\boldmath$\partial$}%
%EndExpansion
}^{2}A_{p}={%
%TCIMACRO{\TeXButton{dirac}{\mbox{\boldmath$\partial$}}}%
%BeginExpansion
\mbox{\boldmath$\partial$}%
%EndExpansion
\cdot%
%TCIMACRO{\TeXButton{dirac}{\mbox{\boldmath$\partial$}}}%
%BeginExpansion
\mbox{\boldmath$\partial$}%
%EndExpansion
}A_{p}+{%
%TCIMACRO{\TeXButton{dirac}{\mbox{\boldmath$\partial$}}}%
%BeginExpansion
\mbox{\boldmath$\partial$}%
%EndExpansion
\wedge%
%TCIMACRO{\TeXButton{dirac}{\mbox{\boldmath$\partial$}}}%
%BeginExpansion
\mbox{\boldmath$\partial$}%
%EndExpansion
}A_{p}\label{14}%
\end{equation}
where ${%
%TCIMACRO{\TeXButton{dirac}{\mbox{\boldmath$\partial$}}}%
%BeginExpansion
\mbox{\boldmath$\partial$}%
%EndExpansion
\cdot%
%TCIMACRO{\TeXButton{dirac}{\mbox{\boldmath$\partial$}}}%
%BeginExpansion
\mbox{\boldmath$\partial$}%
%EndExpansion
}$ is called the \textit{covariant D'Alembertian} and ${%
%TCIMACRO{\TeXButton{dirac}{\mbox{\boldmath$\partial$}}}%
%BeginExpansion
\mbox{\boldmath$\partial$}%
%EndExpansion
\wedge%
%TCIMACRO{\TeXButton{dirac}{\mbox{\boldmath$\partial$}}}%
%BeginExpansion
\mbox{\boldmath$\partial$}%
%EndExpansion
}$ is called the Ricci operator.\footnote{For more details concerning the
square of Dirac (and spin-Dirac operators) on a general Riemann-Cartan
spacetime, see \cite{rodnotte}} If $A_{p}=\frac{1}{p!}A_{\mu_{1}...\mu_{p}}%
%TCIMACRO{\TeXButton{bg}{\mbox{\boldmath{$\gamma$}}}}%
%BeginExpansion
\mbox{\boldmath{$\gamma$}}%
%EndExpansion
^{\mu_{_{1}}}\wedge...\wedge%
%TCIMACRO{\TeXButton{bg}{\mbox{\boldmath{$\gamma$}}}}%
%BeginExpansion
\mbox{\boldmath{$\gamma$}}%
%EndExpansion
^{\mu_{p}}$, we have
\begin{equation}
{%
%TCIMACRO{\TeXButton{dirac}{\mbox{\boldmath$\partial$}}}%
%BeginExpansion
\mbox{\boldmath$\partial$}%
%EndExpansion
\cdot%
%TCIMACRO{\TeXButton{dirac}{\mbox{\boldmath$\partial$}}}%
%BeginExpansion
\mbox{\boldmath$\partial$}%
%EndExpansion
}A_{p}=g^{\alpha\beta}(D_{\partial_{\alpha}}D_{\partial_{\beta}}%
-\Gamma_{\alpha\beta}^{\rho}D_{\partial_{\rho}})A_{p}=\frac{1}{p!}%
g^{\alpha\beta}D_{\alpha}D_{\beta}A_{\alpha_{1}\ldots\alpha_{p}}%
%TCIMACRO{\TeXButton{bg}{\mbox{\boldmath{$\gamma$}}}}%
%BeginExpansion
\mbox{\boldmath{$\gamma$}}%
%EndExpansion
^{\alpha_{1}}\wedge\ldots\wedge%
%TCIMACRO{\TeXButton{bg}{\mbox{\boldmath{$\gamma$}}}}%
%BeginExpansion
\mbox{\boldmath{$\gamma$}}%
%EndExpansion
^{\alpha_{p}},\label{15}%
\end{equation}
Also for ${%
%TCIMACRO{\TeXButton{dirac}{\mbox{\boldmath$\partial$}}}%
%BeginExpansion
\mbox{\boldmath$\partial$}%
%EndExpansion
\wedge%
%TCIMACRO{\TeXButton{dirac}{\mbox{\boldmath$\partial$}}}%
%BeginExpansion
\mbox{\boldmath$\partial$}%
%EndExpansion
}$ in an arbitrary basis (coordinate or orthonormal)%
\begin{equation}
{%
%TCIMACRO{\TeXButton{dirac}{\mbox{\boldmath$\partial$}}}%
%BeginExpansion
\mbox{\boldmath$\partial$}%
%EndExpansion
\wedge%
%TCIMACRO{\TeXButton{dirac}{\mbox{\boldmath$\partial$}}}%
%BeginExpansion
\mbox{\boldmath$\partial$}%
%EndExpansion
}A_{p}=\frac{1}{2}\vartheta^{\alpha}\wedge\vartheta^{\beta}([D_{e_{\alpha}%
},D_{e_{\beta}}]-(L_{\alpha\beta}^{\rho}-L_{\beta\alpha}^{\rho})D_{e_{\rho}%
})A_{p}.\label{16}%
\end{equation}

In particular we show \cite{rodoliv2007}) now that%
\begin{equation}
{%
%TCIMACRO{\TeXButton{dirac}{\mbox{\boldmath$\partial$}}}%
%BeginExpansion
\mbox{\boldmath$\partial$}%
%EndExpansion
\wedge%
%TCIMACRO{\TeXButton{dirac}{\mbox{\boldmath$\partial$}}}%
%BeginExpansion
\mbox{\boldmath$\partial$}%
%EndExpansion
}\vartheta^{\mu}=\mathcal{R}^{\mu},\label{166}%
\end{equation}
where $\mathcal{R}^{\mu}=R_{\nu}^{\mu}\vartheta^{\nu}\in\sec\bigwedge
^{1}T^{\ast}M\hookrightarrow\sec\mathcal{C}\!\ell(M,\mathtt{g})$ are the Ricci
$1$-form fields, such that if $R_{\nu\;\ \sigma\mu}^{\;\mu}$ are the
components of the Riemann tensor we use the convention that $R_{\nu\sigma
}=R_{\nu\;\ \sigma\mu}^{\;\mu}$ are the components of the Ricci tensor.

Applying this operator to the 1-forms of the a $1$-form of the basis
$\{\vartheta^{\mu}\}$, we get:
\begin{equation}
(%
%TCIMACRO{\TeXButton{sd}{\bpartial}}%
%BeginExpansion
\bpartial
%EndExpansion
\wedge%
%TCIMACRO{\TeXButton{sd}{\bpartial}}%
%BeginExpansion
\bpartial
%EndExpansion
)\vartheta^{\mu}=-\frac{1}{2}\text{ }R_{\rho}{}^{\mu}{}_{\alpha\beta
}(\vartheta^{\alpha}\wedge\vartheta^{\beta})\vartheta^{\rho}=-\mathcal{R}%
_{\rho}^{\mu}\vartheta_{\rho},
\end{equation}
Then,
\[
\mathcal{R}_{\rho}^{\mu}\text{ }\vartheta^{\rho}=\mathcal{R}_{\rho}^{\mu
}\llcorner%
%TCIMACRO{\TeXButton{bt}{\mbox{\boldmath{$\theta$}}}}%
%BeginExpansion
\mbox{\boldmath{$\theta$}}%
%EndExpansion
^{\rho}+\mathcal{R}_{\rho}^{\mu}\wedge%
%TCIMACRO{\TeXButton{bt}{\mbox{\boldmath{$\theta$}}}}%
%BeginExpansion
\mbox{\boldmath{$\theta$}}%
%EndExpansion
^{\rho}.
\]
The second term in the r.h.s. of this equation is identically null due the
first Bianchi identity. Then
\begin{align}
\mathcal{R}_{\rho}^{\mu}\llcorner\vartheta^{\rho} &  =\vartheta^{\rho
}\lrcorner\mathcal{R}_{\rho}^{\mu}=-\frac{1}{2}\vartheta^{\rho}\lrcorner
(R_{\rho}{}^{\mu}{}_{\alpha\beta}^{\alpha}\vartheta\wedge\vartheta^{\beta
})\nonumber\\
&  =\frac{1}{2}R_{\rho}{}^{\mu}{}_{\alpha\beta}(g^{\rho\alpha}\vartheta
^{\beta}-g^{\rho\beta}\vartheta^{\alpha})\nonumber\\
&  =g^{\rho\alpha}R_{\rho}{}^{\mu}{}_{\alpha\beta}\text{ }\vartheta^{\beta
}=R_{\beta}^{\mu}\text{ }\vartheta^{\beta}=\mathcal{R}^{\mu},
\end{align}
and Eq.(\ref{166}) is proved.

We next show that for every $A\in\sec\bigwedge^{1}T^{\ast}M\hookrightarrow
\sec\mathcal{C}\!\ell(M,\mathtt{g})$%
\[%
%TCIMACRO{\TeXButton{sd}{\bpartial}}%
%BeginExpansion
\bpartial
%EndExpansion
\wedge%
%TCIMACRO{\TeXButton{sd}{\bpartial}}%
%BeginExpansion
\bpartial
%EndExpansion
A=A_{\mu}%
%TCIMACRO{\TeXButton{sd}{\bpartial}}%
%BeginExpansion
\bpartial
%EndExpansion
\wedge%
%TCIMACRO{\TeXButton{sd}{\bpartial}}%
%BeginExpansion
\bpartial
%EndExpansion
\vartheta^{\mu}.
\]
Indeed, using Eq.(\ref{16}) we can write (using for simplicity a
\textit{coordinate} basis) :%
\begin{align}
&
%TCIMACRO{\TeXButton{sd}{\bpartial}}%
%BeginExpansion
\bpartial
%EndExpansion
\wedge%
%TCIMACRO{\TeXButton{sd}{\bpartial}}%
%BeginExpansion
\bpartial
%EndExpansion
A\nonumber\\
&  =\frac{1}{2}%
%TCIMACRO{\TeXButton{bg}{\mbox{\boldmath{$\gamma$}}}}%
%BeginExpansion
\mbox{\boldmath{$\gamma$}}%
%EndExpansion
^{\alpha}\wedge%
%TCIMACRO{\TeXButton{bg}{\mbox{\boldmath{$\gamma$}}}}%
%BeginExpansion
\mbox{\boldmath{$\gamma$}}%
%EndExpansion
^{\beta}\left\{  [\partial_{\alpha},\partial_{\beta}](A_{\kappa}%
)-\Gamma_{\alpha\beta}^{\rho}\partial_{\rho}(A_{\kappa})+\Gamma_{\beta\alpha
}^{\rho}\partial_{\rho}(A_{\kappa})\right\}
%TCIMACRO{\TeXButton{bg}{\mbox{\boldmath{$\gamma$}}}}%
%BeginExpansion
\mbox{\boldmath{$\gamma$}}%
%EndExpansion
^{\kappa}+A_{\mu}%
%TCIMACRO{\TeXButton{sd}{\bpartial}}%
%BeginExpansion
\bpartial
%EndExpansion
\wedge%
%TCIMACRO{\TeXButton{sd}{\bpartial}}%
%BeginExpansion
\bpartial
%EndExpansion%
%TCIMACRO{\TeXButton{bg}{\mbox{\boldmath{$\gamma$}}}}%
%BeginExpansion
\mbox{\boldmath{$\gamma$}}%
%EndExpansion
^{\mu}\nonumber\\
&  =A_{\mu}%
%TCIMACRO{\TeXButton{sd}{\bpartial}}%
%BeginExpansion
\bpartial
%EndExpansion
\wedge%
%TCIMACRO{\TeXButton{sd}{\bpartial}}%
%BeginExpansion
\bpartial
%EndExpansion%
%TCIMACRO{\TeXButton{bg}{\mbox{\boldmath{$\gamma$}}}}%
%BeginExpansion
\mbox{\boldmath{$\gamma$}}%
%EndExpansion
^{\mu}.\label{16a}%
\end{align}

\end{document}